\documentclass[english,prl, twocolumn, notitlepage, superscriptaddress]{revtex4-1}
\usepackage[T1]{fontenc}
\usepackage{verbatim}
\usepackage{amsmath}
\usepackage{graphicx}
\makeatletter
\usepackage{amsmath}
\usepackage{subfigure}
\usepackage{graphicx, mathtools}
\usepackage[colorlinks=true,linkcolor=MidnightBlue,urlcolor=Black,citecolor=MidnightBlue,anchorcolor=MidnightBlue]{hyperref}\usepackage[dvipsnames]{xcolor}

\makeatother

\usepackage{babel}
\begin{document}
\title{\textcolor{black}{
Controlled optofluidic crystallization of colloids tethered at interfaces
}}
\author{Alessio Caciagli }
\affiliation{Cavendish Laboratory, 19 JJ Thomson Avenue, Cambridge CB3 0HE, UK}
\author{Rajesh Singh }
\affiliation{DAMTP, Centre for Mathematical Sciences, University of Cambridge,
Wilberforce Road, Cambridge CB3 0WA, UK}
\author{Darshana Joshi}
\affiliation{Cavendish Laboratory, 19 JJ Thomson Avenue, Cambridge CB3 0HE, UK}
\author{R. Adhikari}
\affiliation{DAMTP, Centre for Mathematical Sciences, University of Cambridge,
Wilberforce Road, Cambridge CB3 0WA, UK}
\affiliation{The Institute of Mathematical Sciences-HBNI, CIT Campus, Chennai 600113,
India}
\author{Erika Eiser}
\email{ee247@cam.ac.uk}

\affiliation{Cavendish Laboratory, 19 JJ Thomson Avenue, Cambridge CB3 0HE, UK}
\begin{abstract}
We report experiments that show rapid crystallization of colloids tethered to an oil-water interface in response to laser illumination. This light-induced transition is due to a combination of long-ranged thermophoretic pumping and local optical binding.
We show that the flow-induced force on the colloids can be described as the gradient of a potential. 
The non-equilibrium steady state due to local heating thus admits an effective equilibrium description.
The optofluidic manipulation explored in this work opens novel ways to manipulate and assemble colloidal particles.%
\end{abstract}
\maketitle
Since their introduction \cite{Ashkin1986}, optical tweezers have
revolutionised the manipulation of matter at the nano- to micro-meter
scale \cite{Svoboda1993,hofkens1997molecular,Molloy2003,Polin2006,Mizuno2007}.
Tweezers have found extensive use in the trapping and assembly of
micron-sized colloidal particles \cite{Chapin2006,Cizmr2010,Tkalec2011}
and have enabled the formation of  novel forms of colloidal matter that are held together
by optical forces \cite{Burns1990,Vossen2004,Mellor2006,Work2015,Kudo2016,Lin2017,Liu2017}.
While the various mechanisms of optical trapping  are well understood,  much less is known about 
the effects of light-induced heating on the flow of the surrounding fluid medium, in particular near a liquid-liquid interface.
Naively, one might think that, at a liquid-liquid interface, the fluid motion would be dominated by Marangoni flow, which should drive the fluid {\em away} from the hot spot~ \cite{namura2015photothermally,karbalaei2016thermocapillarity}. However, in our experiments, surface tethered colloids moved {\em towards} the hot spot, even when they are out of range of the direct optical binding forces \cite{Wei2016}.

In this Letter, we show that the optical trapping of a single 
colloidal particle near a water-oil interface can set up a long-ranged,
non-equilibrium force field, which causes colloidal particles that are tethered to the surface but otherwise freely diffusing,
to move to the hot spot where they crystallize.
The sign, magnitude, and distance-dependence of this non-equlibrium force cannot be accounted
for by static colloidal interactions nor by surface-tension driven Marangoni flows. 
Using theory and simulation, we show that the experimentally observed 
colloidal motion is produced by stalled thermophoresis
of a single, optically trapped colloid. In fact, to  a good approximation, the stalled particle acts 
as a hydrodynamic monopole. For motion
parallel to the interface, the force is the gradient of a potential
and the particle dynamics admits an effective equilibrium description.
Brownian dynamics simulations in this emergent potential, whose strength
is determined by the local heating, is in excellent agreement with
experiments.

Importantly, the effective attractive potential depends only on the thermophoretic mobility of the stalled colloid. 
It manifests itself in simple solvents such as water or oil and is of much longer range than the
optical trapping potential. As a consequence, interfacially trapped colloids can be used as switchable pumps. 
Such addressable pumps would  enable novel strategies for optofluidic manipulation and the controlled 
assembly of colloidal particles~\cite{Whitesides2002}. 
Below, we describe our experimental results, theoretical analyses, and numerical simulations.
\begin{figure*}
\centering\includegraphics[width=0.99\textwidth]{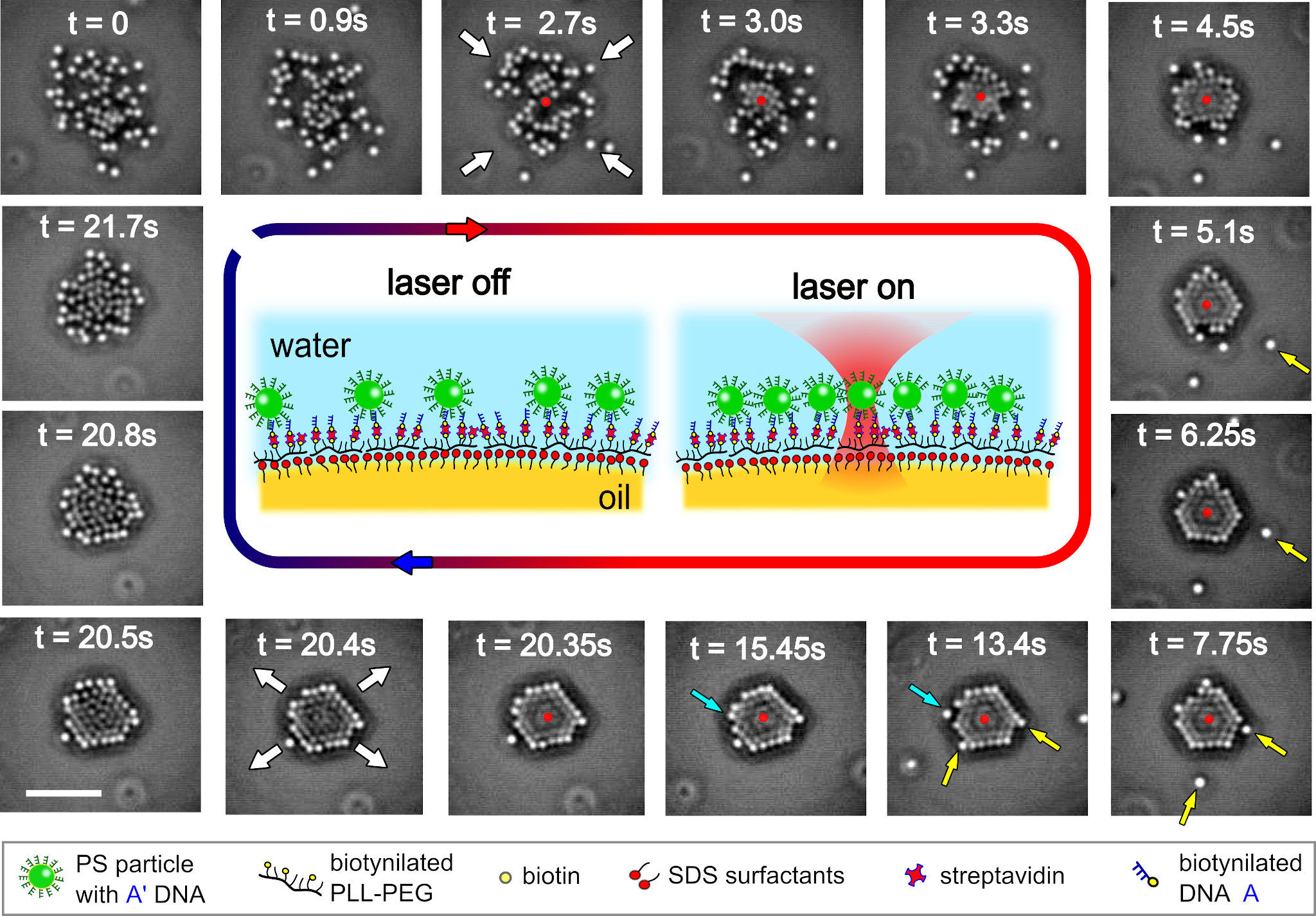}\caption{
The panels in the rim of the figure show a time trace (clockwise) of the video-microscopy images  of the light-induced entrapment and release of  0.53 $\mathrm\mu$m large polystyrene colloids  tethered to the water-oil interface (the sale bar is 20 $\mathrm\mu$m). At $t$=2.7 s, the laser is switched on (red arrow), trapping a single colloid (red dot. The width of the dot indicates the approximate width of the laser focus). Subsequently, opto-phoretic pumping brings other colloids to the hot spot, where they form a crystal stabilised by optical binding. At $t$=20.4 s, the laser is switched off (blue arrow), and the crystal dissolves.  The motion of the particles from yellow and cyan arrows provides clear evidence
of a long-ranged attraction to the center. 
The two central panels  give a schematic side view of the water-oil interface coated  with anionic surfactant (sodium dodecyl sulphate (SDS)) onto which a cationic block-copolymer  (biotinilated polylysine (PLL)-polyethylene glycol (PEG)) is adsorbed. The biotynilated PEG ends are functionalized by streptavidin~\cite{Joshi2016}, which in turn binds to a single-stranded DNA sequence (denoted by A). This A DNA strand then hybridizes with complementary (A') strands on the polystyrene (PS) colloids The keys at the bottom mark each component.  
The particles diffuse freely on the interface when the laser is off (left panel); a single particle is optically trapped when the laser
is turned on (right panel). For details, see movie SV (SI). \label{fig:1}}
\end{figure*}

\emph{Sample geometry:} The inner panels of Fig.\,(\ref{fig:1}) show a sketch of our sample geometry.
Oil  droplets with a radius   between 20$\mathrm{\mu m}$ and 30$\mathrm{\mu m}$ were coated
with a surfactant-polymer layer, following the protocol described in ref.~\cite{Joshi2016}. 
Onto this layer, we grafted a dense brush of single-stranded (ss)DNA sequences (denoted by $A$).
Polystyrene particles (PS) of radius $a=0.53\mathrm{\mu m}$,
functionalised by complementary $A'$ ssDNA strands, were then allowed to hybridize with the $A$ chains on the surface. 
The DNA coating of the colloids prevents
them from aggregating. As the colloids are tethered to the surface, rather than embedded in it,
they are not subject to capillary forces, light-wave reflections or long-ranged
electrostatic dipolar interactions that would be caused by the asymmetry of charge distributions
on interfacially wetted colloids \cite{Park2008,Park2011}. 
Moreover, the colloids do not deform the surfactant-polymer coating, as the DNA tethering keep the colloids some 50 $nm$ away from the surface layer. As the oil droplets are much larger than the colloids, the interface is effectively flat on the scale probed in our experiments. Although tethered to surface polymers, the colloids are otherwise free to diffuse along  the interface. 
 
When  a laser beam is focused above the oil-water surface, it will trap a single tethered colloid. This colloid will act as the thermophoretic ``pump'' that will recruit other tethered colloids. However, a fraction of the colloids remain untethered and diffuse freely in the bulk. These particles serve as tracers of the bulk fluid flow. Further details of the system and the calibration of the trap are provided in the SI \cite{siText}. 

\emph{Reversible crystallization:} The outer panels of Fig.\,(\ref{fig:1})
show our principal experimental result. Following the frames clockwise starting from the top left, the pictures show the crystallisation and dissolution of a colloidal crystal, as the laser us switched on (red arrow) and then switched off (blue arrow).
 The primary optically trapped colloid is shown in red. The
first two frames show free diffusion when the laser is off. When the
laser is turned on in the third frame, diffusion is immediately replaced
by directed motion towards the trapped particle with speeds of upto
$50\,\mathrm{\mu m/s}$. This leads to rapid crystallization which
is essentially complete within a few seconds, as shown in in the frame
at $6.2\mathrm{s}$. Thermal fluctuations cause small displacements
in the core of the crystallite but large ones at the edges where particle
rearrangements take place, as shown in the frames between $6.2\mathrm{s}$
and $20.3\mathrm{s}$. The crystallite begins to melt as soon as the
laser is turned off and a freely diffusing state is recovered within
a few seconds. This cycle of freezing and melting in response to turning
the laser on and off is rapid, robust and reproducible. 

\begin{figure}
\centering\includegraphics[width=0.99\columnwidth]{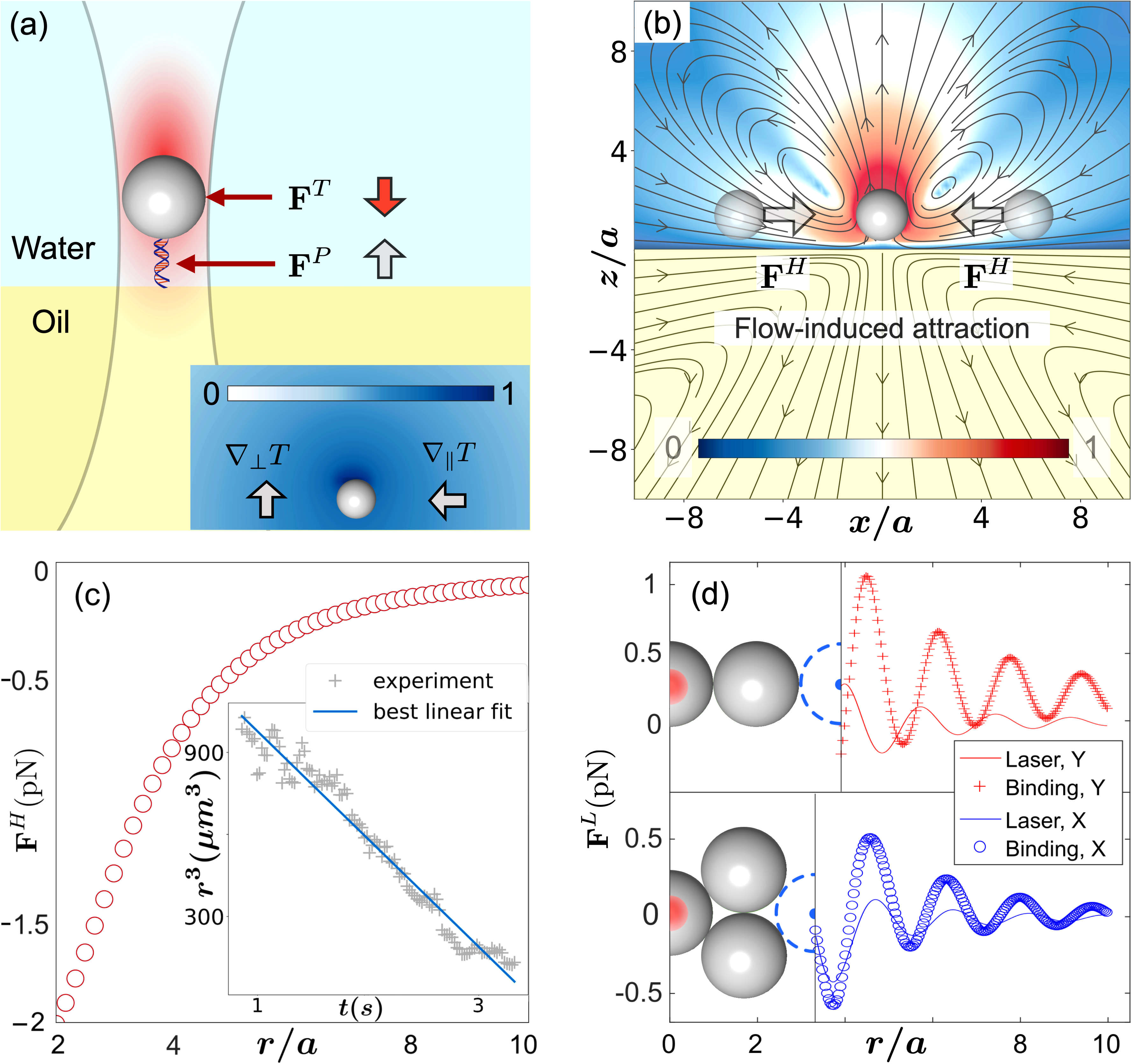}\caption{
\textcolor{black}{Long-ranged optofluidic and short-ranged optical binding forces.} (a)
schematic of a trapped DNA-tethered colloid. The thermophoretic force
$\mathbf{F}^{T}$ drives the colloid towards the interface compressing
the tether and generating a reaction force $\mathbf{F}^{P}$. Motion
stalls when $\mathbf{F}^{T}+\mathbf{F}^{P}=0$. The inset, false color
plot of the temperature, shows gradients perpendicular ($\perp)$
and parallel ($\parallel)$ to the interface. (b) flow streamlines
in the oil and water phases due to the stalled colloid which, to leading
order, is a monopole of strength $\mathbf{-F}^{T}$. The colormap
is the logarithm of the speed of flow in the water phase normalized
by the maximum. The flow entrains tethered untrapped particles leading
to a optofluidic force $\mathbf{F}^{H}$. (c) Inverse-square variation
of the magnitude of $\mathbf{F}^{H}$ with distance $r$ from the
stalled colloid. The inset shows experimental $r(t)$ data for aggregating
particles, following a $r^{3}\sim t$ scaling. (d) The variation of
optical binding forces $\mathbf{F}^{L}$ with normalized $r/a$ parallel
and perpendicular to the polarization of the trapping laser. Solid
and broken lines are contributions from the trapping laser and the
particle scattering forces respectively. The local minima of $\mathbf{F}^{L}$
promote crystalline order. \label{fig:2}}
\end{figure}

\begin{figure}
\centering\includegraphics[width=0.96\columnwidth]{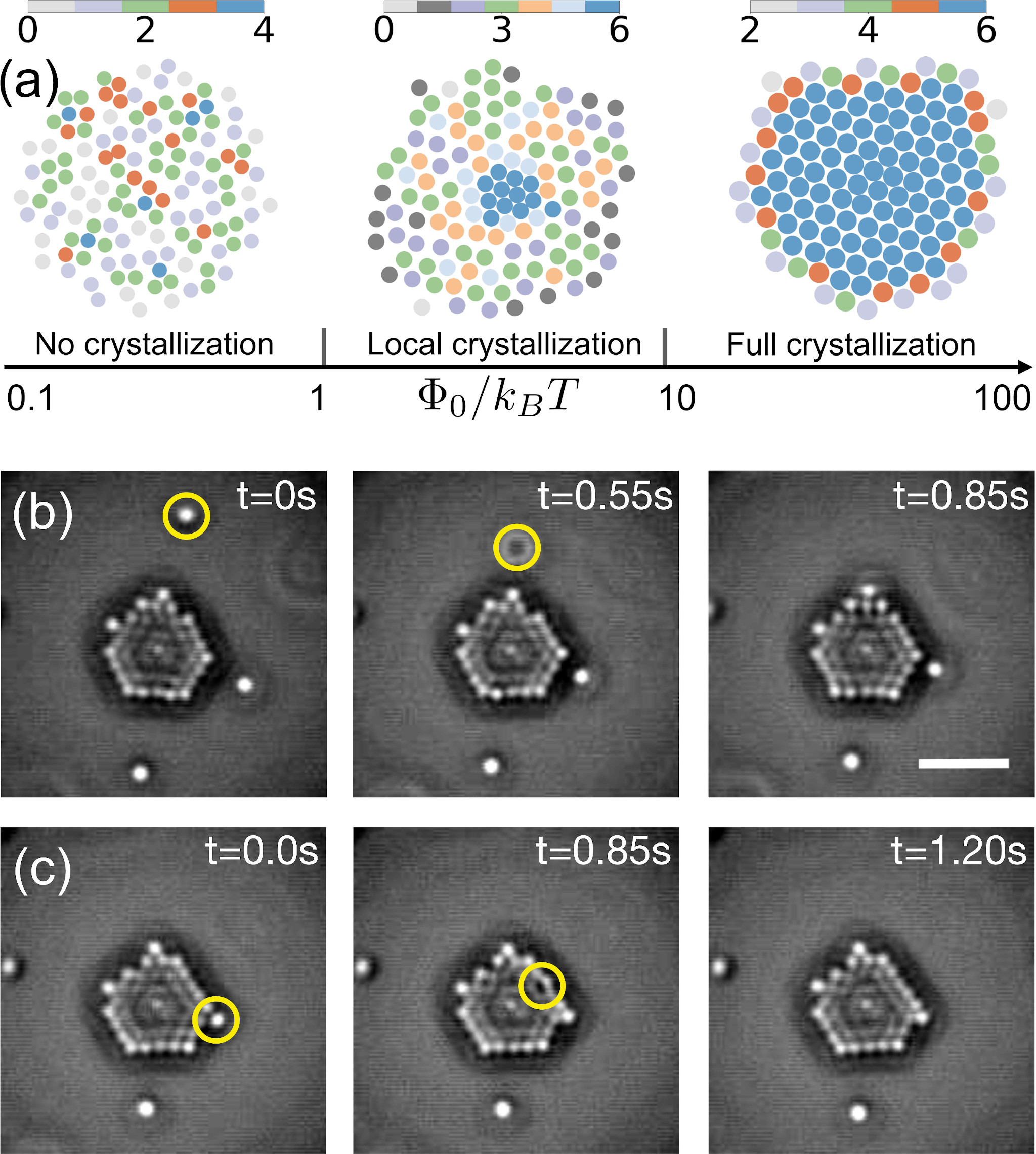}\caption{(a) Competition between the optofluidic potential and Brownian motion
determines the extent of crystalline order. Order is complete when
the strength of the potential $\Phi_{0}/k_{B}T>10$, partial when
$\Phi_{0}/k_{B}T\sim1$, and absent when $\Phi_{0}/k_{B}T<1$. Particles
are coloured by their coordination number, which serves as a measure
of order (b) An untethered particle (yellow circle) is convected away
from the interface by the optofluidic flow of Fig.\,\ref{fig:2}(b).
(c) Short-range forces promote the rapid annealing of a defect (yellow
circle) produced by a particle colliding with the crystal (yellow
circle) in the first frame. The scale bar is $10\mathrm{\mu m}$.
\label{fig:3}}
\end{figure}

\emph{Optofluidic mechanism: }What force underlies this phenomenon?
By examining the motion of the colloids as the laser is turned on,
it is clear that the force has a range of at least $5\mu m$ and a
magnitude of the order of pN, directed radially \emph{inwards} to
the trapped particle. At such distances, neither the direct optical trapping, nor the optical binding forces to be discussed below, can play a role. 
The  entrainment of colloids by Marangoni flow can be immediately ruled out, as such a flow must point \emph{outwards }from the hot region surrounding the
laser focus. Direct thermophoresis toward the hot colloid is also incompatible with the experimental data, as untethered colloids are seen to move first towards the heated colloid and then to move vertically away (Fig.\,\ref{fig:3}(b)):  thermophoresis would result in isotropic attraction. 
 In the absence of other plausible mechanisms, we are
led to postulate the following: the colloid nearest to the laser focus
is optically trapped and local heating near an interface between two fluids with different thermal conductivities induces an assymmetric thermal gradient
in the surrounding fluid. This gradient pushes the colloid towards the interface, where it stalls (as its Soret coefficient is positive \cite{Burelbach2017}). From that moment on, the thermal gradient along its surface drives a thermo-osmotic flow originating in a thin boundary layer around the colloid~\cite{Anderson1989, burelbach2018, burelbach2018unified, wei2020}.  Since the colloid remains stalled, the thermo-osmotic flow continues unabated, but produces no particle motion. This leads to a monopolar
hydrodynamic counterflow in the fluid, with the monopole pointing
normal to the interface and into the water phase. The long-ranged
and attractive character of the flow entrains untrapped
particles and draws them towards the focus. If the entrained colloids are tethered, they  aggregate into crystallites under the action of the
optofluidic force. The local crystalline order is  enhanced by short-ranged
forces, including those due to optical binding (see below).  In contrast, untethered tracer colloids first move along the surface towards the trapped colloid, but then they are advected away from the surface. Such behavior is illustrated  in Fig.\,\ref{fig:3}(b) where an
untethered particle appears and then disappears from the focal plane,
consistent with the flow pattern shown in Fig.\,\ref{fig:2}(b).

To make the above hypothesis quantitative and testable, we solve the
equations of mass, momentum and energy conservation in the fluid with
appropriate boundary conditions at the colloid surfaces and the oil-water
interface (detailed in \cite{siText}). The geometry is shown schematically
in Fig.\,\ref{fig:2}(a). We use the boundary integral representation
for the momentum (Stokes) and energy (Laplace) equations to impose
boundary conditions at the colloid surfaces and use appropriate Green's
functions to satisfy the boundary conditions at the oil-water interface
\cite{singh2019competing}. The integral equations are solved in a
basis of irreducible tensorial harmonics to yield the temperature
field $T$ and fluid flow velocity $\boldsymbol{v}$ in an externally
imposed temperature field $T^{\infty}$ (representing laser heating).
These are shown in Fig.\,\ref{fig:2}(b) for a single trapped colloid.
From these we obtain the thermophoretic force $\mathbf{F}^{T}$ on
the trapped colloid and the optofluidic force $\mathbf{F}^{H}$ with
which free colloids are attracted to the trapped colloid as, 
\begin{alignat}{1}
\mathbf{F}^{T}=-\frac{\mu_{T}}{\mu_{\perp}}\boldsymbol{\nabla}T^{{\scriptscriptstyle \infty}}\big|_{1},\quad\mathbf{F}^{H}=\frac{\mu_{T}}{\mu_{\perp}\mu_{\parallel}}\,\mathbf{G}^{\text{w}}\cdot\boldsymbol{\nabla}T^{{\scriptscriptstyle \infty}}\big|_{1}.
\end{alignat}
In the above, $\mu_{T}$ is the thermophoretic mobility, $\mu_{\perp}$
and $\mu_{\parallel}$ are, respectively, the mobility perpendicular
and parallel to the interface, $\mathbf{G}^{\text{w}}$ is a Green's
function of the Stokes equation for a no-shear plane interface and
$|_{1}$ indicates evaluation at the center of the trapped colloid
\cite{siText}. 
The optofluidic force, through its dependence on $\mathbf{G}^{\text{w}}$,
varies monotonically as the inverse square of the distance $r$ from
the trapped colloid. As particle motion is overdamped, the velocity
scales as $v\negthinspace\sim\negthinspace r^{-2}$, and hence displacements
scale with time intervals as $r^{3}(0)-r^{3}(t)\negmedspace\sim\negmedspace t$.
We test this from the experimentally measured positions to find excellent
agreement, shown in the inset of Fig.\,\ref{fig:2}(c). 

\emph{Optofluidic potential:} For motion at a constant height, the
nonequilibrium optofluidic force $\mathbf{F}^{H}\negthinspace=\negthinspace F^{H}\hat{\boldsymbol{r}}$
admits a potential
\begin{equation}
\Phi(r)=-\frac{\mu_{T}}{4\pi\eta\mu_{\perp}\mu_{\parallel}}\left[\frac{1}{1+\lambda}\frac{h}{r^{*}}+\frac{2\lambda}{1+\lambda}\frac{h^{3}}{r^{*^{3}}}\right]\partial_{z}T^{{\scriptscriptstyle \infty}}.
\end{equation}
Here $r^{*}\negmedspace=\negmedspace\sqrt{r^{2}+h^{2}}$, $h$ is
the height of the colloid from the interface and $\lambda=\eta_{\text{o}}/\eta_{\text{w}}$
is the ratio of the viscosities of oil and water. 
\textcolor{black}{The optofluidic potential depends linearly on the ratio 
$\kappa_{\text{o}}/\kappa_{\text{w}}$ of thermal conductivity of the oil and water layers through the dependence on the temperature gradient \cite{siText}.}
Then, the in-plane
coordinate $\boldsymbol{R}_{i}\negmedspace=\negmedspace(X_{i},Y_{i})$
of an untrapped colloid ($i\negmedspace=\negmedspace2,3,\ldots N)$
obeys the overdamped Langevin equation
\begin{equation}
d\boldsymbol{R}_{i}=-\mu_{\parallel}\boldsymbol{\nabla}_{i}\left(U+\Phi\right)dt+\sqrt{2k_{B}T\mu_{\parallel}}\,d\boldsymbol{\xi}_{i}\label{eq:ito}
\end{equation}
where $U$ is a potential containing the sum of all short-ranged colloid-tether
and colloid-colloid interactions, $\Phi$ is the optofluidic potential
evaluated at the location of the particle, and $d\boldsymbol{\xi}_{i}$
is a zero-mean Gaussian random variable with variance $\langle d\boldsymbol{\xi}_{i}d\boldsymbol{\xi}_{j}\rangle\negthinspace=\negthinspace\delta_{ij}\boldsymbol{I}dt$.
The stationary distribution of the particle positions is Gibbsian,
$P\sim\exp[-(U+\Phi)/k_{B}T]$, even though the dynamics is out of
equilibrium. At an air-water interface, where $\lambda\negthinspace=\negthinspace0$,
the optofluidic potential has a Coulomb form $\Phi\sim1/r^{*}$. The
opposite limit of $\lambda\negthinspace\rightarrow\negthinspace\infty$,
corresponding to a no-slip wall, gives an optofluidic potential $\Phi\sim1/r^{*3}.$
The latter form, with different prefactors, has been found in previous
studies on charged \cite{squires2001effective}, thermophoretic \cite{di2009colloidal,weinert2008observation},
and active colloids \cite{Singh2016,singh2019competing,bolitho2020}. 
\textcolor{black}{Thus, near a liquid-solid boundary, the scaling of Fig.\ref{fig:2}(b) 
is modified to $r^5\propto t$. 
Thus, the optofluidic mechanism described by the monopole has a wider applicability. 
We believe that monopolar flow rationalises a great variety of phenomena in phoretic \cite{Wirnsberger4911} and active matter \cite{bolitho2020} and that its relevance will be widely appreciated in due course.
}

The strength of the potential $\Phi$ when compared with the thermal
energy $k_{B}T$ determines the onset of crystalline order. Denoting
it by
\begin{equation}
\Phi_{0}=\frac{1}{4\pi\eta\mu_{\parallel}(1+\lambda)}\cdot\frac{\mu_{T}}{\mu_{\perp}}\cdot\partial_{z}T^{{\scriptscriptstyle \infty}}
\end{equation}
and using parameters $\mu_{T}\negmedspace=\negmedspace10\mathrm{\mu m^{2}s^{-1}K^{-1}}$,
$\lambda\negmedspace=\negmedspace30$, and $\eta_{\text{w}}\negmedspace=\negmedspace8.9\times10^{-4}\mathrm{Pas}$,
we get $\Phi_{0}\negmedspace\sim\negmedspace100k_{B}T$ when $\partial_{z}T^{{\scriptscriptstyle \infty}}\negmedspace=\negmedspace15K\mu m^{-1}$.
Here we have used the experimentally measured positions in Fig.\,\ref{fig:2}(c)
to estimate the $\partial_{z}T^{{\scriptscriptstyle \infty}}$ from
other known parameters. The strength is  proportional to the thermal
gradient and leads, curiously, to freezing by heating and melting
by cooling. We show this explicitly in Fig\emph{.}\,\ref{fig:3}(a)
by direct numerical simulations of Eq.(\ref{eq:ito}) as a function
of the strength of the optofluidic potential. 

\emph{Short-range forces}: 
Once the long-ranged optofluidic interaction draws particles into
the center of force, short-ranged optical binding forces act to enhance
crystalline order. The optical binding force is obtained from a numerical
solution of the Maxwell equations in the Mie approximation (detailed
in \cite{siText}). It is shorter in range
than the optofluidic force but, being oscillatory and anisotropic
(see Fig.\,\ref{fig:2}(d)), couples to both positional and bond
order of the colloidal crystal. Its effect can be inferred indirectly
from the rapid annealing of a defect (yellow circle) produced by the
collision of a tethered particle with the crystallite, as shown in
Fig.\,\ref{fig:3}(c). We have not studied this coupling in detail
and leave it to future work. 

\emph{Conclusion}: Our experiments shows how a novel non-equilibrium
optofluidic force can be used to transport particles towards (or away)
from an optically trapped ``seed'' particle.
It is important to distinguish that the optofluidic force field is qualitatively different from the light-controlled thermoelectric 
fields  generated in a medium, which contains a mixture of surfactant,
ions, and micellar depletants~\cite{Lin2017}. It is also different from the thermo-osmotic flow generated by the differential heating of trapped Janus particles~\cite{mousavi2019clustering}.
Theoretical analysis
shows that the optofluidic force can be described in terms of the gradient of a potential, whose
strength is proportional to the temperature gradient at the location
of the seed. Untrapped particles couple to this potential regardless
of their material properties, enabling the optofluidic manipulation
of particles that cannot, otherwise, be optically trapped. Both the
location of the potential and its strength can be modulated by the
laser and its sign can be altered by changing the ratio of thermal
conductivities of the liquids. We foresee this to lead to novel mechanisms
of switchable, addressable transport in microfluidics,
\textcolor{black}{controlled self-assembly of active colloids and the meta-material synthesis.
}

    %

\begin{acknowledgments}
We thank Professors M. E. Cates, D. Frenkel, and E. J. Hinch for helpful
discussions. A.C. thanks the ETN-COLLDENSE (H2020-MCSA-ITN-2014, grant
no. 642774) and the Winton Programme for the Physics of Sustainability.
R.S. acknowledges the support of a Royal Society-SERB Newton International
Fellowship. D.J. thanks the Udayan Care-VCare grant, the Nehru Trust
for Cambridge University, the Schlumberger Foundation, Faculty for
the Future Program, and Hughes Hall Santander Bursary Scholarship.
R.A. thanks the Isaac Newton Trust for an Early Career Grant. 
Work was funded in part by the European Research Council under the EU's
Horizon 2020 Program, Grant
No. 740269.
\end{acknowledgments}


\begin{thebibliography}{37}%
\makeatletter
\providecommand \@ifxundefined [1]{%
 \@ifx{#1\undefined}
}%
\providecommand \@ifnum [1]{%
 \ifnum #1\expandafter \@firstoftwo
 \else \expandafter \@secondoftwo
 \fi
}%
\providecommand \@ifx [1]{%
 \ifx #1\expandafter \@firstoftwo
 \else \expandafter \@secondoftwo
 \fi
}%
\providecommand \natexlab [1]{#1}%
\providecommand \enquote  [1]{``#1''}%
\providecommand \bibnamefont  [1]{#1}%
\providecommand \bibfnamefont [1]{#1}%
\providecommand \citenamefont [1]{#1}%
\providecommand \href@noop [0]{\@secondoftwo}%
\providecommand \href [0]{\begingroup \@sanitize@url \@href}%
\providecommand \@href[1]{\@@startlink{#1}\@@href}%
\providecommand \@@href[1]{\endgroup#1\@@endlink}%
\providecommand \@sanitize@url [0]{\catcode `\\12\catcode `\$12\catcode
  `\&12\catcode `\#12\catcode `\^12\catcode `\_12\catcode `\%12\relax}%
\providecommand \@@startlink[1]{}%
\providecommand \@@endlink[0]{}%
\providecommand \url  [0]{\begingroup\@sanitize@url \@url }%
\providecommand \@url [1]{\endgroup\@href {#1}{\urlprefix }}%
\providecommand \urlprefix  [0]{URL }%
\providecommand \Eprint [0]{\href }%
\providecommand \doibase [0]{http://dx.doi.org/}%
\providecommand \selectlanguage [0]{\@gobble}%
\providecommand \bibinfo  [0]{\@secondoftwo}%
\providecommand \bibfield  [0]{\@secondoftwo}%
\providecommand \translation [1]{[#1]}%
\providecommand \BibitemOpen [0]{}%
\providecommand \bibitemStop [0]{}%
\providecommand \bibitemNoStop [0]{.\EOS\space}%
\providecommand \EOS [0]{\spacefactor3000\relax}%
\providecommand \BibitemShut  [1]{\csname bibitem#1\endcsname}%
\let\auto@bib@innerbib\@empty
\bibitem [{\citenamefont {Ashkin}\ \emph {et~al.}(1986)\citenamefont {Ashkin},
  \citenamefont {Dziedzic}, \citenamefont {Bjorkholm},\ and\ \citenamefont
  {Chu}}]{Ashkin1986}%
  \BibitemOpen
  \bibfield  {author} {\bibinfo {author} {\bibfnamefont {A.}~\bibnamefont
  {Ashkin}}, \bibinfo {author} {\bibfnamefont {J.~M.}\ \bibnamefont
  {Dziedzic}}, \bibinfo {author} {\bibfnamefont {J.~E.}\ \bibnamefont
  {Bjorkholm}}, \ and\ \bibinfo {author} {\bibfnamefont {S.}~\bibnamefont
  {Chu}},\ }\href {\doibase 10.1364/OL.11.000288} {\bibfield  {journal}
  {\bibinfo  {journal} {Optics Letters}\ }\textbf {\bibinfo {volume} {11}},\
  \bibinfo {pages} {288} (\bibinfo {year} {1986})},\ \Eprint
  {http://arxiv.org/abs/1411.1912} {1411.1912} \BibitemShut {NoStop}%
\bibitem [{\citenamefont {Svoboda}\ \emph {et~al.}(1993)\citenamefont
  {Svoboda}, \citenamefont {Schmidt}, \citenamefont {Schnapp},\ and\
  \citenamefont {Block}}]{Svoboda1993}%
  \BibitemOpen
  \bibfield  {author} {\bibinfo {author} {\bibfnamefont {K.}~\bibnamefont
  {Svoboda}}, \bibinfo {author} {\bibfnamefont {C.~F.}\ \bibnamefont
  {Schmidt}}, \bibinfo {author} {\bibfnamefont {B.~J.}\ \bibnamefont
  {Schnapp}}, \ and\ \bibinfo {author} {\bibfnamefont {S.~M.}\ \bibnamefont
  {Block}},\ }\href {\doibase 10.1038/365721a0} {\bibfield  {journal} {\bibinfo
   {journal} {Nature}\ }\textbf {\bibinfo {volume} {365}},\ \bibinfo {pages}
  {721} (\bibinfo {year} {1993})}\BibitemShut {NoStop}%
\bibitem [{\citenamefont {Hofkens}\ \emph {et~al.}(1997)\citenamefont
  {Hofkens}, \citenamefont {Hotta}, \citenamefont {Sasaki}, \citenamefont
  {Masuhara},\ and\ \citenamefont {Iwai}}]{hofkens1997molecular}%
  \BibitemOpen
  \bibfield  {author} {\bibinfo {author} {\bibfnamefont {J.}~\bibnamefont
  {Hofkens}}, \bibinfo {author} {\bibfnamefont {J.}~\bibnamefont {Hotta}},
  \bibinfo {author} {\bibfnamefont {K.}~\bibnamefont {Sasaki}}, \bibinfo
  {author} {\bibfnamefont {H.}~\bibnamefont {Masuhara}}, \ and\ \bibinfo
  {author} {\bibfnamefont {K.}~\bibnamefont {Iwai}},\ }\href@noop {} {\bibfield
   {journal} {\bibinfo  {journal} {Langmuir}\ }\textbf {\bibinfo {volume}
  {13}},\ \bibinfo {pages} {414} (\bibinfo {year} {1997})}\BibitemShut
  {NoStop}%
\bibitem [{\citenamefont {Molloy}(2003)}]{Molloy2003}%
  \BibitemOpen
  \bibfield  {author} {\bibinfo {author} {\bibfnamefont {J.~E.}\ \bibnamefont
  {Molloy}},\ }\href {\doibase 10.1126/science.1087148} {\bibfield  {journal}
  {\bibinfo  {journal} {Science}\ }\textbf {\bibinfo {volume} {300}},\ \bibinfo
  {pages} {2045} (\bibinfo {year} {2003})}\BibitemShut {NoStop}%
\bibitem [{\citenamefont {Polin}\ \emph {et~al.}(2006)\citenamefont {Polin},
  \citenamefont {Grier},\ and\ \citenamefont {Quake}}]{Polin2006}%
  \BibitemOpen
  \bibfield  {author} {\bibinfo {author} {\bibfnamefont {M.}~\bibnamefont
  {Polin}}, \bibinfo {author} {\bibfnamefont {D.~G.}\ \bibnamefont {Grier}}, \
  and\ \bibinfo {author} {\bibfnamefont {S.~R.}\ \bibnamefont {Quake}},\ }\href
  {\doibase 10.1103/PhysRevLett.96.088101} {\bibfield  {journal} {\bibinfo
  {journal} {Physical Review Letters}\ }\textbf {\bibinfo {volume} {96}}
  (\bibinfo {year} {2006}),\ 10.1103/PhysRevLett.96.088101}\BibitemShut
  {NoStop}%
\bibitem [{\citenamefont {Mizuno}\ \emph {et~al.}(2007)\citenamefont {Mizuno},
  \citenamefont {Tardin}, \citenamefont {Schmidt},\ and\ \citenamefont
  {MacKintosh}}]{Mizuno2007}%
  \BibitemOpen
  \bibfield  {author} {\bibinfo {author} {\bibfnamefont {D.}~\bibnamefont
  {Mizuno}}, \bibinfo {author} {\bibfnamefont {C.}~\bibnamefont {Tardin}},
  \bibinfo {author} {\bibfnamefont {C.~F.}\ \bibnamefont {Schmidt}}, \ and\
  \bibinfo {author} {\bibfnamefont {F.~C.}\ \bibnamefont {MacKintosh}},\ }\href
  {\doibase 10.1126/science.1134404} {\bibfield  {journal} {\bibinfo  {journal}
  {Science}\ }\textbf {\bibinfo {volume} {315}},\ \bibinfo {pages} {370}
  (\bibinfo {year} {2007})}\BibitemShut {NoStop}%
\bibitem [{\citenamefont {Chapin}\ \emph {et~al.}(2006)\citenamefont {Chapin},
  \citenamefont {Germain},\ and\ \citenamefont {Dufresne}}]{Chapin2006}%
  \BibitemOpen
  \bibfield  {author} {\bibinfo {author} {\bibfnamefont {S.~C.}\ \bibnamefont
  {Chapin}}, \bibinfo {author} {\bibfnamefont {V.}~\bibnamefont {Germain}}, \
  and\ \bibinfo {author} {\bibfnamefont {E.~R.}\ \bibnamefont {Dufresne}},\
  }\href@noop {} {\bibfield  {journal} {\bibinfo  {journal} {Optics express}\
  }\textbf {\bibinfo {volume} {14}},\ \bibinfo {pages} {13095} (\bibinfo {year}
  {2006})}\BibitemShut {NoStop}%
\bibitem [{\citenamefont {{\v{C}}i{\v{z}}m{\'{a}}r}\ \emph
  {et~al.}(2010)\citenamefont {{\v{C}}i{\v{z}}m{\'{a}}r}, \citenamefont
  {Romero}, \citenamefont {Dholakia},\ and\ \citenamefont
  {Andrews}}]{Cizmr2010}%
  \BibitemOpen
  \bibfield  {author} {\bibinfo {author} {\bibfnamefont {T.}~\bibnamefont
  {{\v{C}}i{\v{z}}m{\'{a}}r}}, \bibinfo {author} {\bibfnamefont {L.~C.~D.}\
  \bibnamefont {Romero}}, \bibinfo {author} {\bibfnamefont {K.}~\bibnamefont
  {Dholakia}}, \ and\ \bibinfo {author} {\bibfnamefont {D.~L.}\ \bibnamefont
  {Andrews}},\ }\href@noop {} {\bibfield  {journal} {\bibinfo  {journal}
  {Journal of Physics B: Atomic, Molecular and Optical Physics}\ }\textbf
  {\bibinfo {volume} {43}},\ \bibinfo {pages} {102001} (\bibinfo {year}
  {2010})}\BibitemShut {NoStop}%
\bibitem [{\citenamefont {Tkalec}\ \emph {et~al.}(2011)\citenamefont {Tkalec},
  \citenamefont {Ravnik}, \citenamefont {{\v{C}}opar}, \citenamefont
  {{\v{Z}}umer},\ and\ \citenamefont {Mu{\v{s}}evi{\v{c}}}}]{Tkalec2011}%
  \BibitemOpen
  \bibfield  {author} {\bibinfo {author} {\bibfnamefont {U.}~\bibnamefont
  {Tkalec}}, \bibinfo {author} {\bibfnamefont {M.}~\bibnamefont {Ravnik}},
  \bibinfo {author} {\bibfnamefont {S.}~\bibnamefont {{\v{C}}opar}}, \bibinfo
  {author} {\bibfnamefont {S.}~\bibnamefont {{\v{Z}}umer}}, \ and\ \bibinfo
  {author} {\bibfnamefont {I.}~\bibnamefont {Mu{\v{s}}evi{\v{c}}}},\
  }\href@noop {} {\bibfield  {journal} {\bibinfo  {journal} {Science}\ }\textbf
  {\bibinfo {volume} {333}},\ \bibinfo {pages} {62} (\bibinfo {year}
  {2011})}\BibitemShut {NoStop}%
\bibitem [{\citenamefont {Burns}\ \emph {et~al.}(1990)\citenamefont {Burns},
  \citenamefont {Fournier},\ and\ \citenamefont {Golovchenko}}]{Burns1990}%
  \BibitemOpen
  \bibfield  {author} {\bibinfo {author} {\bibfnamefont {M.~M.}\ \bibnamefont
  {Burns}}, \bibinfo {author} {\bibfnamefont {J.-M.}\ \bibnamefont {Fournier}},
  \ and\ \bibinfo {author} {\bibfnamefont {J.~A.}\ \bibnamefont
  {Golovchenko}},\ }\href {\doibase 10.1126/science.249.4970.749} {\bibfield
  {journal} {\bibinfo  {journal} {Science}\ }\textbf {\bibinfo {volume}
  {249}},\ \bibinfo {pages} {749} (\bibinfo {year} {1990})}\BibitemShut
  {NoStop}%
\bibitem [{\citenamefont {Vossen}\ \emph {et~al.}(2004)\citenamefont {Vossen},
  \citenamefont {Plaisier},\ and\ \citenamefont {van Blaaderen}}]{Vossen2004}%
  \BibitemOpen
  \bibfield  {author} {\bibinfo {author} {\bibfnamefont {D.~L.~J.}\
  \bibnamefont {Vossen}}, \bibinfo {author} {\bibfnamefont {M.~A.}\
  \bibnamefont {Plaisier}}, \ and\ \bibinfo {author} {\bibfnamefont
  {A.}~\bibnamefont {van Blaaderen}},\ }in\ \href {\doibase 10.1117/12.558923}
  {\emph {\bibinfo {booktitle} {Optical Trapping and Optical
  Micromanipulation}}},\ \bibinfo {editor} {edited by\ \bibinfo {editor}
  {\bibfnamefont {K.}~\bibnamefont {Dholakia}}\ and\ \bibinfo {editor}
  {\bibfnamefont {G.~C.}\ \bibnamefont {Spalding}}}\ (\bibinfo {year} {2004})\
  p.\ \bibinfo {pages} {755}\BibitemShut {NoStop}%
\bibitem [{\citenamefont {Mellor}\ and\ \citenamefont
  {Bain}(2006)}]{Mellor2006}%
  \BibitemOpen
  \bibfield  {author} {\bibinfo {author} {\bibfnamefont {C.~D.}\ \bibnamefont
  {Mellor}}\ and\ \bibinfo {author} {\bibfnamefont {C.~D.}\ \bibnamefont
  {Bain}},\ }\href {\doibase 10.1002/cphc.200500348} {\bibfield  {journal}
  {\bibinfo  {journal} {ChemPhysChem}\ }\textbf {\bibinfo {volume} {7}},\
  \bibinfo {pages} {329} (\bibinfo {year} {2006})}\BibitemShut {NoStop}%
\bibitem [{\citenamefont {Work}\ and\ \citenamefont
  {Williams}(2015)}]{Work2015}%
  \BibitemOpen
  \bibfield  {author} {\bibinfo {author} {\bibfnamefont {A.~H.}\ \bibnamefont
  {Work}}\ and\ \bibinfo {author} {\bibfnamefont {S.~J.}\ \bibnamefont
  {Williams}},\ }\href@noop {} {\bibfield  {journal} {\bibinfo  {journal} {Soft
  Matter}\ }\textbf {\bibinfo {volume} {11}},\ \bibinfo {pages} {4266}
  (\bibinfo {year} {2015})}\BibitemShut {NoStop}%
\bibitem [{\citenamefont {Kudo}\ \emph {et~al.}(2016)\citenamefont {Kudo},
  \citenamefont {Wang}, \citenamefont {Yuyama},\ and\ \citenamefont
  {Masuhara}}]{Kudo2016}%
  \BibitemOpen
  \bibfield  {author} {\bibinfo {author} {\bibfnamefont {T.}~\bibnamefont
  {Kudo}}, \bibinfo {author} {\bibfnamefont {S.~F.}\ \bibnamefont {Wang}},
  \bibinfo {author} {\bibfnamefont {K.~I.}\ \bibnamefont {Yuyama}}, \ and\
  \bibinfo {author} {\bibfnamefont {H.}~\bibnamefont {Masuhara}},\ }\href
  {\doibase 10.1021/acs.nanolett.6b00123} {\bibfield  {journal} {\bibinfo
  {journal} {Nano Letters}\ }\textbf {\bibinfo {volume} {16}},\ \bibinfo
  {pages} {3058} (\bibinfo {year} {2016})}\BibitemShut {NoStop}%
\bibitem [{\citenamefont {Lin}\ \emph {et~al.}(2017)\citenamefont {Lin},
  \citenamefont {Zhang}, \citenamefont {Peng}, \citenamefont {Wu},
  \citenamefont {Coughlan}, \citenamefont {Mao}, \citenamefont {Bevan},\ and\
  \citenamefont {Zheng}}]{Lin2017}%
  \BibitemOpen
  \bibfield  {author} {\bibinfo {author} {\bibfnamefont {L.}~\bibnamefont
  {Lin}}, \bibinfo {author} {\bibfnamefont {J.}~\bibnamefont {Zhang}}, \bibinfo
  {author} {\bibfnamefont {X.}~\bibnamefont {Peng}}, \bibinfo {author}
  {\bibfnamefont {Z.}~\bibnamefont {Wu}}, \bibinfo {author} {\bibfnamefont
  {A.~C.~H.}\ \bibnamefont {Coughlan}}, \bibinfo {author} {\bibfnamefont
  {Z.}~\bibnamefont {Mao}}, \bibinfo {author} {\bibfnamefont {M.~A.}\
  \bibnamefont {Bevan}}, \ and\ \bibinfo {author} {\bibfnamefont
  {Y.}~\bibnamefont {Zheng}},\ }\href {\doibase 10.1126/sciadv.1700458}
  {\bibfield  {journal} {\bibinfo  {journal} {Science Advances}\ }\textbf
  {\bibinfo {volume} {3}},\ \bibinfo {pages} {e1700458} (\bibinfo {year}
  {2017})}\BibitemShut {NoStop}%
\bibitem [{\citenamefont {Liu}\ and\ \citenamefont {Li}(2017)}]{Liu2017}%
  \BibitemOpen
  \bibfield  {author} {\bibinfo {author} {\bibfnamefont {J.}~\bibnamefont
  {Liu}}\ and\ \bibinfo {author} {\bibfnamefont {Z.-Y.}\ \bibnamefont {Li}},\
  }\href@noop {} {\bibfield  {journal} {\bibinfo  {journal} {Photonics
  Research}\ }\textbf {\bibinfo {volume} {5}},\ \bibinfo {pages} {201}
  (\bibinfo {year} {2017})}\BibitemShut {NoStop}%
\bibitem [{\citenamefont {Namura}\ \emph {et~al.}(2015)\citenamefont {Namura},
  \citenamefont {Nakajima}, \citenamefont {Kimura},\ and\ \citenamefont
  {Suzuki}}]{namura2015photothermally}%
  \BibitemOpen
  \bibfield  {author} {\bibinfo {author} {\bibfnamefont {K.}~\bibnamefont
  {Namura}}, \bibinfo {author} {\bibfnamefont {K.}~\bibnamefont {Nakajima}},
  \bibinfo {author} {\bibfnamefont {K.}~\bibnamefont {Kimura}}, \ and\ \bibinfo
  {author} {\bibfnamefont {M.}~\bibnamefont {Suzuki}},\ }\href@noop {}
  {\bibfield  {journal} {\bibinfo  {journal} {App. Phys. Lett.}\ }\textbf
  {\bibinfo {volume} {106}},\ \bibinfo {pages} {043101} (\bibinfo {year}
  {2015})}\BibitemShut {NoStop}%
\bibitem [{\citenamefont {Karbalaei}\ \emph {et~al.}(2016)\citenamefont
  {Karbalaei}, \citenamefont {Kumar},\ and\ \citenamefont
  {Cho}}]{karbalaei2016thermocapillarity}%
  \BibitemOpen
  \bibfield  {author} {\bibinfo {author} {\bibfnamefont {A.}~\bibnamefont
  {Karbalaei}}, \bibinfo {author} {\bibfnamefont {R.}~\bibnamefont {Kumar}}, \
  and\ \bibinfo {author} {\bibfnamefont {H.~J.}\ \bibnamefont {Cho}},\
  }\href@noop {} {\bibfield  {journal} {\bibinfo  {journal} {Micromachines}\
  }\textbf {\bibinfo {volume} {7}},\ \bibinfo {pages} {13} (\bibinfo {year}
  {2016})}\BibitemShut {NoStop}%
\bibitem [{\citenamefont {Wei}\ \emph {et~al.}(2016)\citenamefont {Wei},
  \citenamefont {Ng}, \citenamefont {Chan},\ and\ \citenamefont
  {Ou-Yang}}]{Wei2016}%
  \BibitemOpen
  \bibfield  {author} {\bibinfo {author} {\bibfnamefont {M.-T.}\ \bibnamefont
  {Wei}}, \bibinfo {author} {\bibfnamefont {J.}~\bibnamefont {Ng}}, \bibinfo
  {author} {\bibfnamefont {C.~T.}\ \bibnamefont {Chan}}, \ and\ \bibinfo
  {author} {\bibfnamefont {H.~D.}\ \bibnamefont {Ou-Yang}},\ }\href {\doibase
  10.1038/srep38883} {\bibfield  {journal} {\bibinfo  {journal} {Scientific
  Reports}\ }\textbf {\bibinfo {volume} {6}},\ \bibinfo {pages} {38883}
  (\bibinfo {year} {2016})}\BibitemShut {NoStop}%
\bibitem [{\citenamefont {Whitesides}\ and\ \citenamefont
  {Grzybowski}(2002)}]{Whitesides2002}%
  \BibitemOpen
  \bibfield  {author} {\bibinfo {author} {\bibfnamefont {G.}~\bibnamefont
  {Whitesides}}\ and\ \bibinfo {author} {\bibfnamefont {B.}~\bibnamefont
  {Grzybowski}},\ }\href@noop {} {\bibfield  {journal} {\bibinfo  {journal}
  {Science}\ }\textbf {\bibinfo {volume} {295}},\ \bibinfo {pages} {2418}
  (\bibinfo {year} {2002})}\BibitemShut {NoStop}%
\bibitem [{\citenamefont {Joshi}\ \emph {et~al.}(2016)\citenamefont {Joshi},
  \citenamefont {Bargteil}, \citenamefont {Caciagli}, \citenamefont
  {Burelbach}, \citenamefont {Xing}, \citenamefont {Nunes}, \citenamefont
  {Pinto}, \citenamefont {Ara{\'{u}}jo}, \citenamefont {Brujic},\ and\
  \citenamefont {Eiser}}]{Joshi2016}%
  \BibitemOpen
  \bibfield  {author} {\bibinfo {author} {\bibfnamefont {D.}~\bibnamefont
  {Joshi}}, \bibinfo {author} {\bibfnamefont {D.}~\bibnamefont {Bargteil}},
  \bibinfo {author} {\bibfnamefont {A.}~\bibnamefont {Caciagli}}, \bibinfo
  {author} {\bibfnamefont {J.}~\bibnamefont {Burelbach}}, \bibinfo {author}
  {\bibfnamefont {Z.}~\bibnamefont {Xing}}, \bibinfo {author} {\bibfnamefont
  {A.~S.}\ \bibnamefont {Nunes}}, \bibinfo {author} {\bibfnamefont {D.~E.~P.}\
  \bibnamefont {Pinto}}, \bibinfo {author} {\bibfnamefont {N.~A.~M.}\
  \bibnamefont {Ara{\'{u}}jo}}, \bibinfo {author} {\bibfnamefont
  {J.}~\bibnamefont {Brujic}}, \ and\ \bibinfo {author} {\bibfnamefont
  {E.}~\bibnamefont {Eiser}},\ }\href {\doibase 10.1126/sciadv.1600881}
  {\bibfield  {journal} {\bibinfo  {journal} {Science Advances}\ }\textbf
  {\bibinfo {volume} {2}},\ \bibinfo {pages} {e1600881} (\bibinfo {year}
  {2016})}\BibitemShut {NoStop}%
\bibitem [{\citenamefont {Park}\ and\ \citenamefont {Furst}(2008)}]{Park2008}%
  \BibitemOpen
  \bibfield  {author} {\bibinfo {author} {\bibfnamefont {B.~J.}\ \bibnamefont
  {Park}}\ and\ \bibinfo {author} {\bibfnamefont {E.~M.}\ \bibnamefont
  {Furst}},\ }\href {\doibase 10.1021/la802575k} {\bibfield  {journal}
  {\bibinfo  {journal} {Langmuir}\ }\textbf {\bibinfo {volume} {24}},\ \bibinfo
  {pages} {13383} (\bibinfo {year} {2008})}\BibitemShut {NoStop}%
\bibitem [{\citenamefont {Park}\ and\ \citenamefont {Furst}(2011)}]{Park2011}%
  \BibitemOpen
  \bibfield  {author} {\bibinfo {author} {\bibfnamefont {B.~J.}\ \bibnamefont
  {Park}}\ and\ \bibinfo {author} {\bibfnamefont {E.~M.}\ \bibnamefont
  {Furst}},\ }\href {\doibase 10.1039/c1sm00005e} {\bibfield  {journal}
  {\bibinfo  {journal} {Soft Matter}\ }\textbf {\bibinfo {volume} {7}},\
  \bibinfo {pages} {7676} (\bibinfo {year} {2011})}\BibitemShut {NoStop}%
\bibitem [{siT()}]{siText}%
  \BibitemOpen
  \href@noop {} {\enquote {\bibinfo {title} {See supplemental material at [to
  be inserted] which includes the details of the calculations, experiments,
  numerics, and movies of crystallization and includes Refs.[25-49].}}\ }\BibitemShut {NoStop}%
\bibitem [{\citenamefont {Caciagli}\ \emph {et~al.}(2017)\citenamefont
  {Caciagli}, \citenamefont {Joshi}, \citenamefont {Kotar},\ and\ \citenamefont
  {Eiser}}]{caciagli2017optical}%
  \BibitemOpen
  \bibfield  {author} {\bibinfo {author} {\bibfnamefont {A.}~\bibnamefont
  {Caciagli}}, \bibinfo {author} {\bibfnamefont {D.}~\bibnamefont {Joshi}},
  \bibinfo {author} {\bibfnamefont {J.}~\bibnamefont {Kotar}}, \ and\ \bibinfo
  {author} {\bibfnamefont {E.}~\bibnamefont {Eiser}},\ }\href@noop {}
  {\bibfield  {journal} {\bibinfo  {journal} {arXiv preprint arXiv:1703.08210}\
  } (\bibinfo {year} {2017})}\BibitemShut {NoStop}%
\bibitem [{\citenamefont {Yanagishima}\ \emph {et~al.}(2010)\citenamefont
  {Yanagishima}, \citenamefont {Frenkel}, \citenamefont {Kotar},\ and\
  \citenamefont {Eiser}}]{Yanagishima2010}%
  \BibitemOpen
  \bibfield  {author} {\bibinfo {author} {\bibfnamefont {T.}~\bibnamefont
  {Yanagishima}}, \bibinfo {author} {\bibfnamefont {D.}~\bibnamefont
  {Frenkel}}, \bibinfo {author} {\bibfnamefont {J.}~\bibnamefont {Kotar}}, \
  and\ \bibinfo {author} {\bibfnamefont {E.}~\bibnamefont {Eiser}},\ }\href
  {\doibase 10.1088/0953-8984/23/19/194118} {\bibfield  {journal} {\bibinfo
  {journal} {Journal of physics: Condensed matter}\ }\textbf {\bibinfo {volume}
  {23}},\ \bibinfo {pages} {194118} (\bibinfo {year} {2010})},\ \Eprint
  {http://arxiv.org/abs/1010.1211} {arXiv:1010.1211} \BibitemShut {NoStop}%
\bibitem [{\citenamefont {{Di Michele}}\ \emph {et~al.}(2011)\citenamefont {{Di
  Michele}}, \citenamefont {Yanagishima}, \citenamefont {Brewer}, \citenamefont
  {Kotar}, \citenamefont {Eiser},\ and\ \citenamefont
  {Fraden}}]{DiMichele2011}%
  \BibitemOpen
  \bibfield  {author} {\bibinfo {author} {\bibfnamefont {L.}~\bibnamefont {{Di
  Michele}}}, \bibinfo {author} {\bibfnamefont {T.}~\bibnamefont
  {Yanagishima}}, \bibinfo {author} {\bibfnamefont {A.~R.}\ \bibnamefont
  {Brewer}}, \bibinfo {author} {\bibfnamefont {J.}~\bibnamefont {Kotar}},
  \bibinfo {author} {\bibfnamefont {E.}~\bibnamefont {Eiser}}, \ and\ \bibinfo
  {author} {\bibfnamefont {S.}~\bibnamefont {Fraden}},\ }\href {\doibase
  10.1103/PhysRevLett.107.136101} {\bibfield  {journal} {\bibinfo  {journal}
  {Physical Review Letters}\ }\textbf {\bibinfo {volume} {107}} (\bibinfo
  {year} {2011}),\ 10.1103/PhysRevLett.107.136101},\ \Eprint
  {http://arxiv.org/abs/1106.3980} {arXiv:1106.3980} \BibitemShut {NoStop}%
\bibitem [{\citenamefont {Gelles}\ \emph {et~al.}(1988)\citenamefont {Gelles},
  \citenamefont {Schnapp},\ and\ \citenamefont {Sheetz}}]{Gelles1988}%
  \BibitemOpen
  \bibfield  {author} {\bibinfo {author} {\bibfnamefont {J.}~\bibnamefont
  {Gelles}}, \bibinfo {author} {\bibfnamefont {B.~J.}\ \bibnamefont {Schnapp}},
  \ and\ \bibinfo {author} {\bibfnamefont {M.~P.}\ \bibnamefont {Sheetz}},\
  }\href {\doibase 10.1038/331450a0} {\bibfield  {journal} {\bibinfo  {journal}
  {Nature}\ }\textbf {\bibinfo {volume} {331}},\ \bibinfo {pages} {450}
  (\bibinfo {year} {1988})},\ \Eprint {http://arxiv.org/abs/arXiv:1011.1669v3}
  {arXiv:arXiv:1011.1669v3} \BibitemShut {NoStop}%
\bibitem [{\citenamefont {Cheezum}\ \emph {et~al.}(2001)\citenamefont
  {Cheezum}, \citenamefont {Walker},\ and\ \citenamefont
  {Guilford}}]{Cheezum2001}%
  \BibitemOpen
  \bibfield  {author} {\bibinfo {author} {\bibfnamefont {M.~K.}\ \bibnamefont
  {Cheezum}}, \bibinfo {author} {\bibfnamefont {W.~F.}\ \bibnamefont {Walker}},
  \ and\ \bibinfo {author} {\bibfnamefont {W.~H.}\ \bibnamefont {Guilford}},\
  }\href {\doibase 10.1016/S0006-3495(01)75884-5} {\bibfield  {journal}
  {\bibinfo  {journal} {Biophysical journal}\ }\textbf {\bibinfo {volume}
  {81}},\ \bibinfo {pages} {2378} (\bibinfo {year} {2001})}\BibitemShut
  {NoStop}%
\bibitem [{\citenamefont {Otto}\ \emph {et~al.}(2010)\citenamefont {Otto},
  \citenamefont {Czerwinski}, \citenamefont {Gornall}, \citenamefont {Stober},
  \citenamefont {Oddershede}, \citenamefont {Seidel},\ and\ \citenamefont
  {Keyser}}]{Otto2010}%
  \BibitemOpen
  \bibfield  {author} {\bibinfo {author} {\bibfnamefont {O.}~\bibnamefont
  {Otto}}, \bibinfo {author} {\bibfnamefont {F.}~\bibnamefont {Czerwinski}},
  \bibinfo {author} {\bibfnamefont {J.~L.}\ \bibnamefont {Gornall}}, \bibinfo
  {author} {\bibfnamefont {G.}~\bibnamefont {Stober}}, \bibinfo {author}
  {\bibfnamefont {L.~B.}\ \bibnamefont {Oddershede}}, \bibinfo {author}
  {\bibfnamefont {R.}~\bibnamefont {Seidel}}, \ and\ \bibinfo {author}
  {\bibfnamefont {U.~F.}\ \bibnamefont {Keyser}},\ }\href {\doibase
  10.1364/OE.18.022722} {\bibfield  {journal} {\bibinfo  {journal} {Optics
  Express}\ }\textbf {\bibinfo {volume} {18}},\ \bibinfo {pages} {22722}
  (\bibinfo {year} {2010})}\BibitemShut {NoStop}%
\bibitem [{\citenamefont {Hell}\ \emph {et~al.}(1993)\citenamefont {Hell},
  \citenamefont {Reiner}, \citenamefont {Cremer},\ and\ \citenamefont
  {Stelzer}}]{Hell1993}%
  \BibitemOpen
  \bibfield  {author} {\bibinfo {author} {\bibfnamefont {S.~W.}\ \bibnamefont
  {Hell}}, \bibinfo {author} {\bibfnamefont {G.}~\bibnamefont {Reiner}},
  \bibinfo {author} {\bibfnamefont {C.}~\bibnamefont {Cremer}}, \ and\ \bibinfo
  {author} {\bibfnamefont {E.~H.~K.}\ \bibnamefont {Stelzer}},\ }\href@noop {}
  {\bibfield  {journal} {\bibinfo  {journal} {Journal of Microscopy}\ }\textbf
  {\bibinfo {volume} {169}},\ \bibinfo {pages} {391} (\bibinfo {year}
  {1993})}\BibitemShut {NoStop}%
\bibitem [{\citenamefont {Brettschneider}\ \emph {et~al.}(2011)\citenamefont
  {Brettschneider}, \citenamefont {Volpe}, \citenamefont {Helden},
  \citenamefont {Wehr},\ and\ \citenamefont {Bechinger}}]{Brettschneider2011}%
  \BibitemOpen
  \bibfield  {author} {\bibinfo {author} {\bibfnamefont {T.}~\bibnamefont
  {Brettschneider}}, \bibinfo {author} {\bibfnamefont {G.~G.}\ \bibnamefont
  {Volpe}}, \bibinfo {author} {\bibfnamefont {L.}~\bibnamefont {Helden}},
  \bibinfo {author} {\bibfnamefont {J.}~\bibnamefont {Wehr}}, \ and\ \bibinfo
  {author} {\bibfnamefont {C.}~\bibnamefont {Bechinger}},\ }\href {\doibase
  10.1103/PhysRevE.83.041113} {\bibfield  {journal} {\bibinfo  {journal}
  {Physical Review E - Statistical, Nonlinear, and Soft Matter Physics}\
  }\textbf {\bibinfo {volume} {83}},\ \bibinfo {pages} {1} (\bibinfo {year}
  {2011})},\ \Eprint {http://arxiv.org/abs/arXiv:1009.2386v1}
  {arXiv:arXiv:1009.2386v1} \BibitemShut {NoStop}%
\bibitem [{\citenamefont {Jones}\ \emph {et~al.}(2015)\citenamefont {Jones},
  \citenamefont {Marag{\'{o}}},\ and\ \citenamefont {Volpe}}]{Jones2015}%
  \BibitemOpen
  \bibfield  {author} {\bibinfo {author} {\bibfnamefont {P.}~\bibnamefont
  {Jones}}, \bibinfo {author} {\bibfnamefont {O.}~\bibnamefont {Marag{\'{o}}}},
  \ and\ \bibinfo {author} {\bibfnamefont {G.}~\bibnamefont {Volpe}},\
  }\href@noop {} {\emph {\bibinfo {title} {{Optical Tweezers: Principles and
  Applications}}}}\ (\bibinfo  {publisher} {Cambridge University Press},\
  \bibinfo {year} {2015})\BibitemShut {NoStop}%
\bibitem [{\citenamefont {Bera}\ \emph {et~al.}(2017)\citenamefont {Bera},
  \citenamefont {Paul}, \citenamefont {Singh}, \citenamefont {Ghosh},
  \citenamefont {Kundu}, \citenamefont {Banerjee},\ and\ \citenamefont
  {Adhikari}}]{bera2017fast}%
  \BibitemOpen
  \bibfield  {author} {\bibinfo {author} {\bibfnamefont {S.}~\bibnamefont
  {Bera}}, \bibinfo {author} {\bibfnamefont {S.}~\bibnamefont {Paul}}, \bibinfo
  {author} {\bibfnamefont {R.}~\bibnamefont {Singh}}, \bibinfo {author}
  {\bibfnamefont {D.}~\bibnamefont {Ghosh}}, \bibinfo {author} {\bibfnamefont
  {A.}~\bibnamefont {Kundu}}, \bibinfo {author} {\bibfnamefont
  {A.}~\bibnamefont {Banerjee}}, \ and\ \bibinfo {author} {\bibfnamefont
  {R.}~\bibnamefont {Adhikari}},\ }\href {\doibase 10.1038/srep41638}
  {\bibfield  {journal} {\bibinfo  {journal} {Sci. Rep.}\ }\textbf {\bibinfo
  {volume} {7}} (\bibinfo {year} {2017}),\ 10.1038/srep41638}\BibitemShut
  {NoStop}%
\bibitem [{\citenamefont {Singh}\ \emph {et~al.}(2018)\citenamefont {Singh},
  \citenamefont {Ghosh},\ and\ \citenamefont {Adhikari}}]{singh2018fast}%
  \BibitemOpen
  \bibfield  {author} {\bibinfo {author} {\bibfnamefont {R.}~\bibnamefont
  {Singh}}, \bibinfo {author} {\bibfnamefont {D.}~\bibnamefont {Ghosh}}, \ and\
  \bibinfo {author} {\bibfnamefont {R.}~\bibnamefont {Adhikari}},\ }\href
  {\doibase 10.1103/PhysRevE.98.012136} {\bibfield  {journal} {\bibinfo
  {journal} {Phys. Rev. E}\ }\textbf {\bibinfo {volume} {98}},\ \bibinfo
  {pages} {012136} (\bibinfo {year} {2018})}\BibitemShut {NoStop}%
\bibitem [{\citenamefont {Anderson}(1989)}]{Anderson1989}%
  \BibitemOpen
  \bibfield  {author} {\bibinfo {author} {\bibfnamefont {J.}~\bibnamefont
  {Anderson}},\ }\href {\doibase 10.1146/annurev.fluid.21.1.61} {\bibfield
  {journal} {\bibinfo  {journal} {Annual Review of Fluid Mechanics}\ }\textbf
  {\bibinfo {volume} {21}},\ \bibinfo {pages} {61} (\bibinfo {year}
  {1989})}\BibitemShut {NoStop}%
\bibitem [{\citenamefont {Singh}\ and\ \citenamefont
  {Adhikari}(2018)}]{singh2018generalized}%
  \BibitemOpen
  \bibfield  {author} {\bibinfo {author} {\bibfnamefont {R.}~\bibnamefont
  {Singh}}\ and\ \bibinfo {author} {\bibfnamefont {R.}~\bibnamefont
  {Adhikari}},\ }\href {\doibase 10.1088/2399-6528/aaab0d} {\bibfield
  {journal} {\bibinfo  {journal} {J. Phys. Commun.}\ }\textbf {\bibinfo
  {volume} {2}},\ \bibinfo {pages} {25025} (\bibinfo {year}
  {2018})}\BibitemShut {NoStop}%
\bibitem [{\citenamefont {{R. Singh}}\ \emph {et~al.}(2019)\citenamefont {{R.
  Singh}}, \citenamefont {Adhikari},\ and\ \citenamefont
  {Cates}}]{singh2019competing}%
  \BibitemOpen
  \bibfield  {author} {\bibinfo {author} {\bibnamefont {{R. Singh}}}, \bibinfo
  {author} {\bibfnamefont {R.}~\bibnamefont {Adhikari}}, \ and\ \bibinfo
  {author} {\bibfnamefont {M.~E.}\ \bibnamefont {Cates}},\ }\href {\doibase
  10.1063/1.5090179} {\bibfield  {journal} {\bibinfo  {journal} {The Journal of
  Chemical Physics}\ }\textbf {\bibinfo {volume} {151}},\ \bibinfo {pages}
  {044901} (\bibinfo {year} {2019})}\BibitemShut {NoStop}%
\bibitem [{\citenamefont {Aderogba}\ and\ \citenamefont
  {Blake}(1978)}]{aderogba1978action}%
  \BibitemOpen
  \bibfield  {author} {\bibinfo {author} {\bibfnamefont {K.}~\bibnamefont
  {Aderogba}}\ and\ \bibinfo {author} {\bibfnamefont {J.}~\bibnamefont
  {Blake}},\ }\href {\doibase 10.1017/S0004972700008819} {\bibfield  {journal}
  {\bibinfo  {journal} {Bulletin of the Australian Mathematical Society}\
  }\textbf {\bibinfo {volume} {19}},\ \bibinfo {pages} {309} (\bibinfo {year}
  {1978})}\BibitemShut {NoStop}%
\bibitem [{\citenamefont {Pozrikidis}(1992)}]{pozrikidis1992}%
  \BibitemOpen
  \bibfield  {author} {\bibinfo {author} {\bibfnamefont {C.}~\bibnamefont
  {Pozrikidis}},\ }\href {\doibase 10.1017/CBO9780511624124} {\emph {\bibinfo
  {title} {{Boundary Integral and Singularity Methods for Linearized Viscous
  Flow}}}}\ (\bibinfo  {publisher} {Cambridge University Press},\ \bibinfo
  {year} {1992})\BibitemShut {NoStop}%
\bibitem [{\citenamefont {Singh}\ and\ \citenamefont
  {Adhikari}(2020)}]{singh2019hydrodynamic}%
  \BibitemOpen
  \bibfield  {author} {\bibinfo {author} {\bibfnamefont {R.}~\bibnamefont
  {Singh}}\ and\ \bibinfo {author} {\bibfnamefont {R.}~\bibnamefont
  {Adhikari}},\ }\href {\doibase 10.21105/joss.02318} {\bibfield  {journal}
  {\bibinfo  {journal} {Journal of Open Source Software}\ }\textbf {\bibinfo
  {volume} {5}},\ \bibinfo {pages} {2318} (\bibinfo {year} {2020})}\BibitemShut
  {NoStop}%
\bibitem [{\citenamefont {Mandel}\ and\ \citenamefont
  {Wolf}(1995)}]{Mandel1995}%
  \BibitemOpen
  \bibfield  {author} {\bibinfo {author} {\bibfnamefont {L.}~\bibnamefont
  {Mandel}}\ and\ \bibinfo {author} {\bibfnamefont {E.}~\bibnamefont {Wolf}},\
  }\href {\doibase 10.1017/CBO9781139644105} {\emph {\bibinfo {title} {{Optical
  coherence and quantum optics}}}}\ (\bibinfo  {publisher} {Cambridge
  University Press},\ \bibinfo {address} {Cambridge},\ \bibinfo {year}
  {1995})\BibitemShut {NoStop}%
\bibitem [{\citenamefont {Wolf}(1959)}]{Wolf1959}%
  \BibitemOpen
  \bibfield  {author} {\bibinfo {author} {\bibfnamefont {E.}~\bibnamefont
  {Wolf}},\ }\href {\doibase 10.1098/rspa.1959.0199} {\bibfield  {journal}
  {\bibinfo  {journal} {Proceedings of the Royal Society of London. Series A.
  Mathematical and Physical Sciences}\ }\textbf {\bibinfo {volume} {253}},\
  \bibinfo {pages} {349} (\bibinfo {year} {1959})}\BibitemShut {NoStop}%
\bibitem [{\citenamefont {Richards}\ and\ \citenamefont
  {Wolf}(1959)}]{Richards1959}%
  \BibitemOpen
  \bibfield  {author} {\bibinfo {author} {\bibfnamefont {B.}~\bibnamefont
  {Richards}}\ and\ \bibinfo {author} {\bibfnamefont {E.}~\bibnamefont
  {Wolf}},\ }\href {\doibase 10.1098/rspa.1959.0200} {\bibfield  {journal}
  {\bibinfo  {journal} {Proceedings of the Royal Society of London. Series A.
  Mathematical and Physical Sciences}\ } (\bibinfo {year} {1959}),\
  10.1098/rspa.1959.0200}\BibitemShut {NoStop}%
\bibitem [{\citenamefont {{A. Caciagli}}(2020)}]{cacialgli2020}%
  \BibitemOpen
  \bibfield  {author} {\bibinfo {author} {\bibnamefont {{A. Caciagli}}},\
  }\href@noop {} {Ph.D. thesis},\ \bibinfo  {school} {University of Cambridge}
  (\bibinfo {year} {2020})\BibitemShut {NoStop}%
\bibitem [{\citenamefont {Burelbach}\ \emph {et~al.}(2017)\citenamefont
  {Burelbach}, \citenamefont {Zupkauskas}, \citenamefont {Lamboll},
  \citenamefont {Lan},\ and\ \citenamefont {Eiser}}]{Burelbach2017}%
  \BibitemOpen
  \bibfield  {author} {\bibinfo {author} {\bibfnamefont {J.}~\bibnamefont
  {Burelbach}}, \bibinfo {author} {\bibfnamefont {M.}~\bibnamefont
  {Zupkauskas}}, \bibinfo {author} {\bibfnamefont {R.}~\bibnamefont {Lamboll}},
  \bibinfo {author} {\bibfnamefont {Y.}~\bibnamefont {Lan}}, \ and\ \bibinfo
  {author} {\bibfnamefont {E.}~\bibnamefont {Eiser}},\ }\href {\doibase
  10.1063/1.5001023} {\bibfield  {journal} {\bibinfo  {journal} {The Journal of
  Chemical Physics}\ }\textbf {\bibinfo {volume} {147}},\ \bibinfo {pages}
  {094906} (\bibinfo {year} {2017})},\ \Eprint
  {http://arxiv.org/abs/1705.05279} {1705.05279} \BibitemShut {NoStop}%
\bibitem [{\citenamefont {Burelbach}\ \emph
  {et~al.}(2018{\natexlab{a}})\citenamefont {Burelbach}, \citenamefont
  {Brückner}, \citenamefont {Frenkel},\ and\ \citenamefont
  {Eiser}}]{burelbach2018}%
  \BibitemOpen
  \bibfield  {author} {\bibinfo {author} {\bibfnamefont {J.}~\bibnamefont
  {Burelbach}}, \bibinfo {author} {\bibfnamefont {D.~B.}\ \bibnamefont
  {Brückner}}, \bibinfo {author} {\bibfnamefont {D.}~\bibnamefont {Frenkel}},
  \ and\ \bibinfo {author} {\bibfnamefont {E.}~\bibnamefont {Eiser}},\ }\href
  {\doibase 10.1039/C8SM01132J} {\bibfield  {journal} {\bibinfo  {journal}
  {Soft Matter}\ }\textbf {\bibinfo {volume} {14}},\ \bibinfo {pages} {7446}
  (\bibinfo {year} {2018}{\natexlab{a}})}\BibitemShut {NoStop}%
\bibitem [{\citenamefont {Burelbach}\ \emph
  {et~al.}(2018{\natexlab{b}})\citenamefont {Burelbach}, \citenamefont
  {Frenkel}, \citenamefont {Pagonabarraga},\ and\ \citenamefont
  {Eiser}}]{burelbach2018unified}%
  \BibitemOpen
  \bibfield  {author} {\bibinfo {author} {\bibfnamefont {J.}~\bibnamefont
  {Burelbach}}, \bibinfo {author} {\bibfnamefont {D.}~\bibnamefont {Frenkel}},
  \bibinfo {author} {\bibfnamefont {I.}~\bibnamefont {Pagonabarraga}}, \ and\
  \bibinfo {author} {\bibfnamefont {E.}~\bibnamefont {Eiser}},\ }\href
  {\doibase 10.1140/epje/i2018-11610-3} {\bibfield  {journal} {\bibinfo
  {journal} {The European Physical Journal E}\ }\textbf {\bibinfo {volume}
  {41}},\ \bibinfo {pages} {7} (\bibinfo {year}
  {2018}{\natexlab{b}})}\BibitemShut {NoStop}%
\bibitem [{\citenamefont {Wei}\ \emph {et~al.}(2020)\citenamefont {Wei},
  \citenamefont {Ramírez-Hinestrosa}, \citenamefont {Dobnikar},\ and\
  \citenamefont {Frenkel}}]{wei2020}%
  \BibitemOpen
  \bibfield  {author} {\bibinfo {author} {\bibfnamefont {J.}~\bibnamefont
  {Wei}}, \bibinfo {author} {\bibfnamefont {S.}~\bibnamefont
  {Ramírez-Hinestrosa}}, \bibinfo {author} {\bibfnamefont {J.}~\bibnamefont
  {Dobnikar}}, \ and\ \bibinfo {author} {\bibfnamefont {D.}~\bibnamefont
  {Frenkel}},\ }\href {\doibase 10.1039/C9SM02053E} {\bibfield  {journal}
  {\bibinfo  {journal} {Soft Matter}\ }\textbf {\bibinfo {volume} {16}},\
  \bibinfo {pages} {3621} (\bibinfo {year} {2020})}\BibitemShut {NoStop}%
\bibitem [{\citenamefont {Squires}(2001)}]{squires2001effective}%
  \BibitemOpen
  \bibfield  {author} {\bibinfo {author} {\bibfnamefont {T.~M.}\ \bibnamefont
  {Squires}},\ }\href {\doibase 10.1017/S0022112001005432} {\bibfield
  {journal} {\bibinfo  {journal} {Journal of Fluid Mechanics}\ }\textbf
  {\bibinfo {volume} {443}},\ \bibinfo {pages} {403} (\bibinfo {year}
  {2001})}\BibitemShut {NoStop}%
\bibitem [{\citenamefont {{Di Leonardo}}\ \emph {et~al.}(2009)\citenamefont
  {{Di Leonardo}}, \citenamefont {Ianni},\ and\ \citenamefont
  {Ruocco}}]{di2009colloidal}%
  \BibitemOpen
  \bibfield  {author} {\bibinfo {author} {\bibfnamefont {R.}~\bibnamefont {{Di
  Leonardo}}}, \bibinfo {author} {\bibfnamefont {F.}~\bibnamefont {Ianni}}, \
  and\ \bibinfo {author} {\bibfnamefont {G.}~\bibnamefont {Ruocco}},\ }\href
  {\doibase 10.1021/la8038335} {\bibfield  {journal} {\bibinfo  {journal}
  {Langmuir}\ }\textbf {\bibinfo {volume} {25}},\ \bibinfo {pages} {4247}
  (\bibinfo {year} {2009})}\BibitemShut {NoStop}%
\bibitem [{\citenamefont {Weinert}\ and\ \citenamefont
  {Braun}(2008)}]{weinert2008observation}%
  \BibitemOpen
  \bibfield  {author} {\bibinfo {author} {\bibfnamefont {F.~M.}\ \bibnamefont
  {Weinert}}\ and\ \bibinfo {author} {\bibfnamefont {D.}~\bibnamefont
  {Braun}},\ }\href {\doibase 10.1103/PhysRevLett.101.168301} {\bibfield
  {journal} {\bibinfo  {journal} {Physical Review Letters}\ }\textbf {\bibinfo
  {volume} {101}},\ \bibinfo {pages} {168301} (\bibinfo {year}
  {2008})}\BibitemShut {NoStop}%
\bibitem [{\citenamefont {Singh}\ and\ \citenamefont
  {Adhikari}(2016)}]{Singh2016}%
  \BibitemOpen
  \bibfield  {author} {\bibinfo {author} {\bibfnamefont {R.}~\bibnamefont
  {Singh}}\ and\ \bibinfo {author} {\bibfnamefont {R.}~\bibnamefont
  {Adhikari}},\ }\href {\doibase 10.1103/PhysRevLett.117.228002} {\bibfield
  {journal} {\bibinfo  {journal} {Physical Review Letters}\ }\textbf {\bibinfo
  {volume} {117}},\ \bibinfo {pages} {228002} (\bibinfo {year}
  {2016})}\BibitemShut {NoStop}%
\bibitem [{\citenamefont {Bolitho}\ \emph {et~al.}(2020)\citenamefont
  {Bolitho}, \citenamefont {Singh},\ and\ \citenamefont
  {Adhikari}}]{bolitho2020}%
  \BibitemOpen
  \bibfield  {author} {\bibinfo {author} {\bibfnamefont {A.}~\bibnamefont
  {Bolitho}}, \bibinfo {author} {\bibfnamefont {R.}~\bibnamefont {Singh}}, \
  and\ \bibinfo {author} {\bibfnamefont {R.}~\bibnamefont {Adhikari}},\
  }\href@noop {} {\bibfield  {journal} {\bibinfo  {journal} {Physical Review
  Letters}\ }\textbf {\bibinfo {volume} {124}},\ \bibinfo {pages} {088003}
  (\bibinfo {year} {2020})}\BibitemShut {NoStop}%
\bibitem [{\citenamefont {Wirnsberger}\ \emph {et~al.}(2017)\citenamefont
  {Wirnsberger}, \citenamefont {Fijan}, \citenamefont {Lightwood},
  \citenamefont {{\v S}ari{\'c}}, \citenamefont {Dellago},\ and\ \citenamefont
  {Frenkel}}]{Wirnsberger4911}%
  \BibitemOpen
  \bibfield  {author} {\bibinfo {author} {\bibfnamefont {P.}~\bibnamefont
  {Wirnsberger}}, \bibinfo {author} {\bibfnamefont {D.}~\bibnamefont {Fijan}},
  \bibinfo {author} {\bibfnamefont {R.~A.}\ \bibnamefont {Lightwood}}, \bibinfo
  {author} {\bibfnamefont {A.}~\bibnamefont {{\v S}ari{\'c}}}, \bibinfo
  {author} {\bibfnamefont {C.}~\bibnamefont {Dellago}}, \ and\ \bibinfo
  {author} {\bibfnamefont {D.}~\bibnamefont {Frenkel}},\ }\href {\doibase
  10.1073/pnas.1621494114} {\bibfield  {journal} {\bibinfo  {journal}
  {Proceedings of the National Academy of Sciences}\ }\textbf {\bibinfo
  {volume} {114}},\ \bibinfo {pages} {4911} (\bibinfo {year}
  {2017})}\BibitemShut {NoStop}%
\bibitem [{\citenamefont {Mousavi}\ \emph {et~al.}(2019)\citenamefont
  {Mousavi}, \citenamefont {Kasianiuk}, \citenamefont {Kasyanyuk},
  \citenamefont {Velu}, \citenamefont {Callegari}, \citenamefont
  {Biancofiore},\ and\ \citenamefont {Volpe}}]{mousavi2019clustering}%
  \BibitemOpen
  \bibfield  {author} {\bibinfo {author} {\bibfnamefont {S.~M.}\ \bibnamefont
  {Mousavi}}, \bibinfo {author} {\bibfnamefont {I.}~\bibnamefont {Kasianiuk}},
  \bibinfo {author} {\bibfnamefont {D.}~\bibnamefont {Kasyanyuk}}, \bibinfo
  {author} {\bibfnamefont {S.~K.}\ \bibnamefont {Velu}}, \bibinfo {author}
  {\bibfnamefont {A.}~\bibnamefont {Callegari}}, \bibinfo {author}
  {\bibfnamefont {L.}~\bibnamefont {Biancofiore}}, \ and\ \bibinfo {author}
  {\bibfnamefont {G.}~\bibnamefont {Volpe}},\ }\href@noop {} {\bibfield
  {journal} {\bibinfo  {journal} {Soft matter}\ }\textbf {\bibinfo {volume}
  {15}},\ \bibinfo {pages} {5748} (\bibinfo {year} {2019})}\BibitemShut
  {NoStop}%
\end{thebibliography}
\end{document}